\documentclass[12pt]{elsarticle}
\makeatletter
\def\ps@pprintTitle{%
 \let\@oddhead\@empty
 \let\@evenhead\@empty
 \def\@oddfoot{\centerline{\thepage}}%
 \let\@evenfoot\@oddfoot}
\makeatother

\def\be{\begin{equation}}
\def\ee{\end{equation}}

\usepackage{amsmath}
\usepackage{amssymb}
\usepackage{mathtools}
\usepackage{cases}
\usepackage{bm}
\usepackage{epsfig}
\usepackage{graphicx}
\usepackage{color}
\usepackage{exscale}
\usepackage{relsize}
\usepackage{slantsc}
\usepackage{tikz}

\begin{document}

\begin{frontmatter}



\title{High-Temperature Quantum Coherence from Dissipative Environments\\ }


\author{George E. Cragg}


\address{Statoil, 2103 Citywest Blvd., Suite 800, Houston, TX 77042, USA \\

}

\ead{cragg@alum.mit.edu}

\begin{abstract}
The Feynman-Vernon path integral formalism is used to derive the density matrix of a quantum oscillator that is linearly coupled to an environmental reservoir.  Although low-temperature reservoirs thermalize the oscillator to the usual Boltzmann distribution, reservoirs at intermediate temperatures reduce this distribution to a single, coherent ground state.  Associated with this state is an imaginary frequency indicating an environment which absorbs energy from the oscillator through the suppression of all excited modes.  Further increase of the environmental temperature results again in the thermalization of the quantum oscillator to the expected Boltzmann distribution.  Qualitatively, this result could account for high-temperature quantum effects including the superconducting properties of graphite grains as well as the quantum coherence observed in photosynthetic systems.
\end{abstract}


\end{frontmatter}

\setcounter{footnote}{0}

\section{Introduction}
Quantum coherence is a phenomenon that could have profound impact on technological development.  Usually understood to appear in temperature regimes much lower than those encountered in everyday practice, quantum coherence effects are not directly utilized in the operation of most conventional devices.  Superconducting sensors, magnets and transmission cables are examples where macroscopic quantum coherence is maintained at relatively high temperature.  Nonetheless, even the highest-$T_c$ materials currently remain in the realm of cryogenic temperatures, thus significantly hampering their potential applicability.  Consequently, much effort has been expended in the quest for higher temperature materials, as room-temperature superconductivity remains one of the most prominent issues in condensed matter physics.  Although recent experimental evidence may suggest that superconductivity is possible at room temperature \cite{Esquinazi}, what is perhaps even more compelling is the emerging evidence that photosynthetic systems found in nature may have already evolved a means of exploiting quantum coherence effects at physiological temperatures \cite{EngelFleming, PanitEngel}.  Due to the broad array of organisms sustained by trapping then converting sunlight into chemical energy, photosynthesis is probably the cleanest, most efficient energy collection mechanism on earth, perhaps making high-temperature quantum coherence far more ubiquitous than previously thought. 

Central to the understanding of temperature-dependent quantum coherence is the decoherence or thermalization of quantum systems with their surrounding environment.  In the well-known work of Caldeira and Leggett \cite{CaldeiraLeggett}, the Feynman-Vernon influence functional approach \cite{FeynmanVernon} is used to model a thermalizing environment with the purpose of introducing decoherence into a quantum oscillator.  Following the same path integral approach, it is demonstrated that a thermalizing environment need not always lead to decoherence in the steady-state.  Arriving at this result requires the realization that because the reservoir cannot excite energies too far in excess of its mean thermal energy, a temperature cutoff must be introduced.  As a result, low-temperature environments are found to thermalize the quantum oscillator to the usual Boltzmann distribution of energy levels, whereas higher-temperature regimes induce a single, coherent oscillator ground state with an imaginary frequency.  Since the concomitant imaginary energy eigenvalues correspond to decay rates, all excited states become more unstable with increasing quantum number.  In this way, environments in this higher-temperature regime damp out the excitations by extracting energy from the quantum oscillator, forcing the oscillator to remain in the stable, coherent ground state.  Increasing to very high environmental temperatures once again thermalizes the quantum oscillator in the usual way.  

In short, a dissipative environment can cause a suppression or damping of excited modes, thus acting to maintain quantum coherence even at relatively high temperatures.  This result may underpin recent observations of room-temperature superconducting effects in water-treated graphite grains \cite{Esquinazi}.  Finally, given their emergence and subsequent disappearance of quantum coherence with increasing temperature, photosynthetic molecules most remarkably follow the qualitative behavior described here \cite{EngelFleming, PanitEngel}.
\section{Influence Functional Approach: General Considerations}
\begin{figure} [h]
\begin{center}
\epsfxsize=3.25in\epsfbox{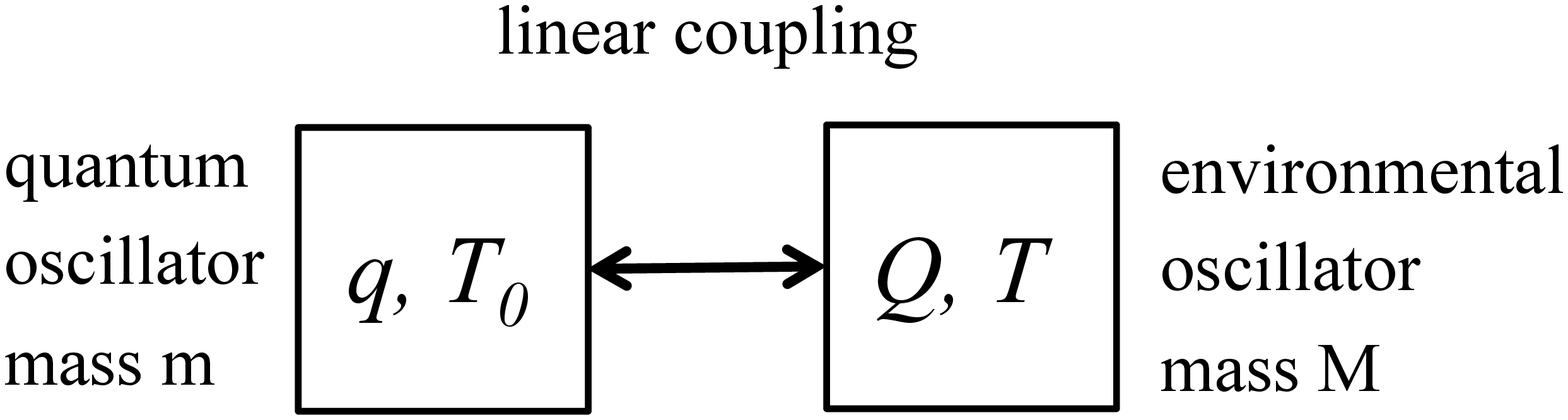}
\vspace{-1.4in}
\end{center}
\caption {A schematic showing the quantum oscillator of interest with coordinates $q$, linearly coupled to a single environmental oscillator with coordinates $Q$.  Initially, there is a temperature difference between the two as oscillator $q$ is at temperature $T_0$ whereas the environment is at $T$. \label{Qq}}
\end{figure}

Depicted schematically in Fig.\,\ref{Qq}, a quantum harmonic oscillator, initially at some temperature $T_0$, linearly coupled to an environmental reservoir.  Ultimately the environment will be modeled as a collection of oscillators at temperature $T$, but we first consider the case of a single-oscillator reservoir, carrying out integrals over a continuum later.  Acquiring the steady-state of the quantum harmonic oscillator requires the time evolution of its corresponding density matrix.
Oscillator $q$ occupies the usual Boltzmann distribution of eigenstates associated with temperature $T_0$.  Because each state occurs with a relative probability of $e^{-\beta_0 E_{n'}}$, the initial density matrix is given by
\begin{subequations}
	\begin{align}
	\rho_0 (q_i, q_i') &= \sum_{n'} e^{-\beta_0 E_{n'}} \phi_{n'} (q_i) \phi_{n'}^\ast (q_i')   
	\label{denmat1} \\
	Z_0 &= Tr[\rho_0] = \int \! \rho_0(q_i, q_i) \, dq_i  ,
	\label{Z01}
	\end{align}
\end{subequations}
\noindent where $\{ \phi_{n'} (q_i) \}$ are the unperturbed eigenfunctions of oscillator $q$, and $q_i$ is shorthand for the initial point $q(0)$ of the oscillator trajectory.  Given a harmonic oscillator of mass $m$ and frequency $\omega_0$, expression \eqref{denmat1} can be summed explicitly to give\footnote{See equation (10.44) of Feynman and Hibbs \cite{FeynmanHibbs}.}
\begin{align}
	\! \! \rho_0(q_i, q_i') &=  \displaystyle \sqrt{\frac{m \omega_0}{2 \pi \hbar 
	\sinh (\beta_0 \hbar \omega_0)}} \exp\left\{ - \frac{m \omega_0}{2 \hbar 
	\sinh (\beta_0 \hbar \omega_0)} \left[ (q_i^2 + q_i^{\prime \, 2} ) 
	\cosh(\beta_0 \hbar \omega_0)
	\right. \right. \nonumber \\
	& \qquad \qquad \qquad \qquad \qquad \qquad \qquad \qquad \qquad
	\qquad \qquad \quad
	\left. \left.  - 2 q_i q_i' \, \right] \vphantom{ \frac{m \, \omega_0}{2 \hbar 
	\sinh (\beta_0 \hbar \omega_0)}} \! \right\} , \label{denmat0}
\end{align}
\noindent and
\be
	Z_0 = \displaystyle \frac{1}{2 \sinh \left( \frac{\beta_0 \hbar \omega_0}{2} \right) } . \label{Z02}
\ee
Throughout, the parameters $\beta$ and $\beta_0$ assume the usual meanings as the respective inverse temperatures,
\be \label{beta0}
	\beta = \frac{1}{k_B T}, \qquad  \beta_0 = \frac{1}{k_B T_0}.
\ee
Alternatively, the initial state of the oscillator $q$ may be taken to be one of the eigenstates $\{ \phi_{n'} (q_i) \}$ so that
\be
	\rho_0(q_i,q_i') = \phi_{n'}(q_i) \phi_{n'}^\ast (q_i') \, , \qquad \, Z_0 = 1 .
\ee
In either case, the time evolution of the density matrix is given by
\be \label{timeev}
	\rho(q_f, q_f',t) = \int e^{ \frac{i}{\hbar} \left[ S(q) - S(q') \right] } F(q,q',t)
	\frac{1}{Z_0} \, \rho_0(q_i,q_i') Dq  Dq'  dq_i  dq_i'  .
\ee
Just as $q_i$ refers to the initial trajectory point $q(0)$, $q_f$ is shorthand for the final point $q(t)$.  Furthermore, the density matrix of the quantum oscillator is expressed as a path integral in which the coordinates of the environmental oscillator $\{ Q \} $ have been integrated out.  As a result, there emerges an influence functional, $F(q,q',t)$, which describes the effect of the environmental oscillator solely in terms of the quantum oscillator coordinates.  Effectively, the coupled problem has been reduced to the dynamics of an isolated oscillator weighted by the kernel $F(q,q',t)$.  Further details may be found in \cite{FeynmanVernon} and \cite{FeynmanHibbs}.  For the case of a thermal reservoir modeled by a harmonic oscillator, the influence functional has the form
\begin{align} \label{F1}
	F(q,q',t) &= \exp \Biggl\{ - \int \limits_0^t \left[ q(t') - q'(t') \right] \int \limits_0^{t'} 
	\left[ q(t'') \alpha (t'-t'') \right. \nonumber \\
	& \qquad \qquad \qquad \qquad \qquad \qquad \qquad
	\left.  
	- q'(t'') \alpha^\ast (t'-t'') \right] dt'' \, dt' \vphantom{- \int \limits_0^t} \Biggr\} ,
\end{align}
where
\be
	\alpha(t'-t'') = \frac{c^2}{2M \omega \hbar} \frac{1}{e^{\beta \hbar \omega} - 1}
	\left[ e^{i \omega (t'-t'')} + e^{\beta \hbar \omega} e^{-i \omega (t'-t'')} \right] .
\ee
These expressions for $F(q,q',t)$ and $\alpha(t'-t'')$ are well-known since they form the basis of the Caldeira-Leggett model \cite{CaldeiraLeggett}, having been directly obtained from Sec.\,12-9 of \cite{FeynmanHibbs}.  

Consider the first integral in the exponential of $F$, $\int_0^{t'} [q(t'') \alpha(t'-t'') - q'(t'') \alpha^\ast (t'-t'')] dt''$.  Because the environment is modeled by a continuum of oscillators, we must sum this expression over all frequencies $\omega_j >0$.  
The contribution from the j{\it th} oscillator is
\begin{align}
	&\int \limits_0^{t'} \left[ q(t'') \alpha_j(t'-t'') - q'(t'') \alpha_j^\ast(t'-t'') \right] \! dt'' = \nonumber \\
	& \qquad \qquad \qquad 
	\frac{c_j^2}{2M_j \omega_j \hbar} \frac{1}{e^{\beta \hbar \omega_j} - 1}
	\int \limits_0^{t'} \left\{ q(t'') \! \left[ e^{i \omega_j (t'-t'')} + e^{\beta \hbar \omega_j}
	e^{-i \omega_j (t'-t'')} \right] \right. \nonumber \\
	& \qquad \qquad \qquad \qquad \qquad \qquad \qquad
	\left. -q'(t'') \! \left[ e^{-i \omega_j (t'-t'')} + e^{\beta \hbar \omega_j} e^{i \omega_j (t'-t'')} \right]
	\right\} \! dt'' . 
\end{align}
Defining the spectral density as
\be \label{Jdef}
	J(\omega) = \pi \sum_j \frac{c_j^2}{2 M_j \omega_j} \delta (\omega - \omega_j) ,
\ee
the sum over all oscillators can be written as an integral
\begin{align}
	\sum_j &\int \limits_0^{t'} \left[ q(t'') \alpha_j(t'-t'') - q'(t'') \alpha_j^\ast(t'-t'') \right] \! dt'' 
	\nonumber \\
	& \qquad  
	= \frac{1}{\pi \hbar} \int \limits_0^\infty \frac{J(\omega)}{e^{\beta \hbar \omega} - 1}
	\int \limits_0^{t'} \left\{ q(t'') \! \left[ e^{i \omega (t'-t'')} + e^{\beta \hbar \omega}
	e^{-i \omega (t'-t'')} \right] \right. \nonumber \\
	& \qquad \qquad \qquad \qquad \quad 
	\left. -q'(t'') \! \left[ e^{-i \omega (t'-t'')} + e^{\beta \hbar \omega} e^{i \omega (t'-t'')} \right]
	\right\} dt'' d \omega\, . 
\end{align}
It is convenient to define
\be \label{adef1}
	a(t'-t'') = \frac{1}{\pi \hbar} \int \limits_0^\infty \frac{J(\omega)}{e^{\beta \hbar \omega} -1}
	\left[ e^{i \omega (t'-t'')} + e^{\beta \hbar \omega} e^{-i \omega (t'-t'')} \right] d \omega \, .
\ee
Assuming the spectral density, $J(\omega)$, to be real, the real part of $a(t'-t'')$ is
\be \label{aRdef1}
	\begin{split}
	Re\left\{ a(t'-t'') \right\} = a_R (t'-t'') &= \frac{1}{\pi \hbar} \int \limits_0^\infty 
	\coth \! \left( \frac{\beta \hbar \omega}{2} \right) \! J(\omega) \cos [\omega (t'-t'')] d \omega \\ 
	&= a_R(t''-t') \, .
	\end{split}
\ee
Similarly, the imaginary part of $a(t'-t'')$ is
\be \label{aIdef1}
	Im \left\{ a(t'-t'') \right\} = a_I(t'-t'') = -\frac{1}{\pi \hbar} \int \limits_0^\infty J(\omega)
	\sin[\omega (t'-t'')] d \omega \, .
\ee
Using the definition of $a(t'-t'')$, the sum over all oscillators simplifies to
\begin{align}
	\sum_j &\int \limits_0^{t'} \left[ q(t'') \alpha_j(t'-t'') - q'(t'') \alpha_j^\ast(t'-t'') \right] \! dt''  
	\nonumber \\
	&= \int \limits_0^{t'} \left\{ a_R(t'-t'') \left[ q(t'') - q'(t'') \right]  + i a_I(t'-t'') \left[ q(t'') + q'(t'') \right]
	\right\} dt''  . 
\end{align}
Substituting this expression into the influence functional \eqref{F1} yields
\allowdisplaybreaks
\begin{align} \label{F2}
	F(q,q',t) &= \exp \Biggl\{ -\frac{1}{2} \int \limits_0^t \left[ q(t') - q'(t') \right] 
	\int \limits_0^t a_R(t'-t'') \left[ q(t'') - q'(t'') \right] dt'' dt'  \nonumber \\
	& \hphantom{\exp \left\{ \right\} }
	 -i \int \limits_0^t \left[ q(t') - q'(t') \right] \int \limits_0^{t'} a_I (t'-t'') \left[
	q(t'') + q'(t'') \right] dt'' dt' \Biggr\} . \nonumber \\[-.05in]
\end{align}

Recall that the action for the quantum oscillator of interest, $q$, is given by
\be
	S(q) = \frac{1}{2} m \! \int \limits_0^t \left[ \dot{q}^2 (t') - \omega_0^2 q^2(t') \right] \! dt' .
\ee
Given this action in addition to result obtained for the influence functional \eqref{F2}, the time evolution of the density matrix is
\begin{align} \label{denmat2}
	Z_0 \, \rho(q_f,q_f',t) &= \int \exp \left( \frac{im}{2 \hbar} \int \limits_0^t \biggl\{ 
	\vphantom{\int \limits_0^t} \dot{q}^2 (t')
	-\omega_0^2 \, q^2(t') -\dot{q}^{\prime \, 2} (t') + \omega_0^2 \, q^{\prime \, 2} (t') 
	\right. \nonumber \\
	& \qquad 
	+ \frac{i \hbar}{m} \left[ q(t') - q'(t') \right] \int \limits_0^t a_R(t'-t'')
	\left[ q(t'') - q'(t'') \right] dt'' \nonumber \\
	& \qquad
	\left.  - \frac{2 \hbar}{m} \left[ q(t') - q'(t') \right] \int \limits_0^{t'} a_I(t'-t'')
	\left[ q(t'') + q'(t'') \right] dt'' \biggr\} dt' \right) \nonumber \\
	& \qquad \qquad \cdot \rho_0 (q_i,q'_i) Dq \, Dq' \, dq_i \, dq'_i .
\end{align}
Transforming to the rotated coordinates,
\begin{subequations} \label{xydef}
	\begin{align}
	x(t') &= \frac{1}{\sqrt{2}} \left[ q(t') + q'(t') \right] \label{xdef} \\[.1in]
	y(t') &= \frac{1}{\sqrt{2}} \left[ q(t') - q'(t') \right] , \label{ydef}
	\end{align}
\end{subequations}
\eqref{denmat2} may be written more succinctly as
\begin{align} \label{denmatxy1}
	\begin{split}
	Z_0 \, &\rho(x_f,y_f,t) = \int \! \exp \left\{ \frac{im}{\hbar} \int \limits_0^t \biggl[  
	\dot{x}(t') \dot{y}(t')
	-\omega_0^{\,2} x(t')y(t') \right. \\
	& \left. + i \frac{\hbar}{m} \, y(t') \int \limits_0^t a_R(t'-t'') y(t'') dt''
	- \frac{2\hbar}{m} \, y(t') \int \limits_0^{t'} a_I (t'-t'') x(t'') dt'' \biggr] dt' \right\} \\
	& \qquad \qquad \qquad \qquad \qquad \qquad
	\cdot \rho_0 (x_i,y_i) Dx \, Dy \, dx_i \, dy_i .
	\end{split}
\end{align} 
As indicated by the resulting expression, the density matrix gives rise to a single, composite action
\begin{align} \label{Action1}
	S[\dot{x}(t'), \dot{y}(t'), x(t'), y(t'), t] &= m\int \limits_0^t \biggl[ \dot{x}(t') \dot{y}(t') 
	-\omega_0^{\,2} \, x(t') y(t') \nonumber \\
	& \qquad \quad
	+ i \frac{\hbar}{m} \, y(t') \! \int \limits_0^t a_R(t'-t'') y(t'') \, dt'' \\
	& \qquad \quad
	- \frac{2 \hbar}{m} \, y(t') \! \int \limits_0^{t'} a_I(t'-t'') x(t'') \, dt'' \biggr] dt' . \nonumber 
\end{align}
Let $x_{cl}(t')$ be the classical trajectory between the fixed initial point $x_i$ and the fixed final point $x_f$, and let $y_{cl}(t')$ be the classical trajectory between the fixed initial point $y_i$ and the fixed final point $y_f$.  By definition, $x_{cl}(t')$ and $y_{cl}(t')$ must extremize the action $S$.  In the usual way, $x(t')$ and $y(t')$ are expressed as the sum of their corresponding extremal paths plus their respective fluctuations, $\delta x(t')$ and $\delta y(t')$:
\setlength{\jot}{.1in}
\begin{subequations}
	\begin{align}
	x(t') &= x_{cl}(t') + \delta x(t') \\
	Dx(t') &= D \delta x(t') \, ,
	\end{align}
\end{subequations}
\begin{subequations}
	\begin{align}
	y(t') &= y_{cl}(t') + \delta y(t') \\
	Dy(t') &= D \delta y(t') \, .
	\end{align}
\end{subequations}
In terms of $x_{cl}(t')$, $\delta x(t')$ and $y_{cl}(t')$, $\delta y(t')$, the composite action becomes
\begin{align} \label{Action3}
	&S[\dot{x}(t'), \dot{y}(t'), x(t'), y(t'), t] = \nonumber \\ 
	&
	m\int \limits_0^t \biggl[ \dot{x}_{cl}(t') \dot{y}_{cl}(t')
	- \ddot{x}_{cl}(t') \delta y(t') - \ddot{y}_{cl} (t') \delta x(t') + \delta \dot{x}(t') \delta \dot{y}(t') 
	- \omega_0^2 \, x_{cl}(t')y_{cl}(t') \nonumber \\
	& \qquad \qquad \qquad \qquad \qquad \quad
	- \omega_0^2 \, x_{cl}(t') \delta y(t') - \omega_0^2 \, y_{cl}(t') \delta x(t') 
	- \omega_0^2 \, \delta x(t') \delta y(t') \nonumber \\
	& \quad 
	+ i \frac{\hbar}{m} \, y_{cl}(t') \! \int \limits_0^t a_R(t'-t'') \, y_{cl}(t'') \, dt'' 
	+ i \frac{2 \hbar}{m} \, \delta y(t') \! \int \limits_0^t a_R (t'-t'') \, y_{cl}(t'') \, dt'' \nonumber \\
	& \quad
	+ i \frac{\hbar}{m} \, \delta y(t') \! \int \limits_0^t a_R(t'-t'') \, \delta y(t'') \, dt'' 
	-\frac{2 \hbar}{m} \, x_{cl}(t') \! \int \limits_{t'}^t a_I(t''-t') \, y_{cl}(t'') \, dt''  \nonumber \\
	& \quad
	-\frac{2 \hbar}{m} \, \delta y(t') \! \int \limits_0^{t'} a_I(t'-t'') \, x_{cl}(t'') \, dt'' 
	-\frac{2 \hbar}{m} \, \delta x(t') \! \int \limits_{t'}^t a_I(t''-t') \, y_{cl}(t'') \, dt'' \nonumber \\
	& \quad
	-\frac{2 \hbar}{m} \, \delta y(t') \! \int \limits_0^{t'} a_I(t'-t'') \, \delta x(t'') \, dt'' \biggr] dt' .
\end{align}
Through the rearrangement of integrals, all terms linear in $\delta x(t'')$ and $\delta y(t'')$ have been expressed in terms of $\delta x(t')$ and $\delta y(t')$.  The linear terms in the fluctuations vanish, thereby constraining the classical trajectories to satisfy
\begin{subequations} \label{classtrajxy}
	\be \label{ycl} 
	\ddot{y}_{cl} (t') + \omega_0^{\,2} \, y_{cl}(t') + \frac{2 \hbar}{m} \! \int \limits_{t'}^t a_I (t''-t') 
	\, y_{cl}(t'') \, dt'' = 0 ,
	\ee \vspace{-.1in} 
	\begin{align} \label{xcl}
	\begin{split}
	\ddot{x}_{cl} (t') + \omega_0^{\,2} \, x_{cl}(t') 
	&- i \frac{2 \hbar}{m} \! \int \limits_0^t a_R(t'-t'') \, y_{cl}(t'') \, dt'' \\[-.1in]
	& \qquad \quad
	+ \frac{2 \hbar}{m} \! \int \limits_0^{t'} a_I (t'-t'') \, x_{cl}(t'') \, dt'' = 0 .
	\end{split}
	\end{align} 
\end{subequations}
\noindent Substitution of the resulting simplified action back into the time evolution of the density matrix \eqref{denmatxy1} gives
\be \label{denmatxy3}
	Z_0 \, \rho(x_f,y_f,t) = \mathfrak{T}(t) \! \int e^{\frac{i}{\hbar} S_{cl}} \rho_0(x_i, y_i) \, 
	dx_i \, dy_i ,
\ee
where the path integral dependent prefactor is identified as
\begin{align} \label{prefac}
	\mathfrak{T}(t) &= \int \limits_0^0 \exp \left\{ \frac{im}{\hbar} \int \limits_0^t 
	\biggl[ \delta \dot{x} (t') \delta \dot{y} (t') - \omega_0^{\,2} \, \delta x(t') \delta y(t') 
	\right.  \nonumber \\
	& \qquad \qquad \qquad \qquad
	+ i \frac{\hbar}{m} \, \delta y (t') \! \int \limits_0^t a_R(t'-t'') \, \delta y(t'') \, dt''  \\
	& \qquad \qquad \qquad \qquad
	\left. \vphantom{\frac{1}{2}} -\frac{2 \hbar}{m} \, \delta y(t') \int \limits_0^{t'} a_I (t'-t'')
	\, \delta x(t'') \, dt'' \biggr] dt' \right\} D\delta x D \delta y ,\nonumber
\end{align}
and where the effective classical action is
\be \label{ClassAction1}
	\begin{split}
	S_{cl} &= m\int \limits_0^t 
	\biggl[ \dot{x}_{cl}(t') \dot{y}_{cl}(t') - \omega_0^{\,2} \, x_{cl}(t')y_{cl}(t') 
	+ i \frac{\hbar}{m} \, y_{cl}(t') \! \int \limits_0^t a_R(t'-t'') \, y_{cl}(t'') \, dt''  \\
	& \qquad \qquad \qquad \qquad \qquad \qquad \qquad
	 -\frac{2 \hbar}{m} \, x_{cl}(t') \int \limits_{t'}^t a_I (t''-t') \,y_{cl}(t'') \, dt'' \biggr] dt' .
	 \end{split}
\ee
From equation \eqref{ycl} for $y_{cl}(t')$, we have $-\omega_0^{\,2} \, x_{cl}(t') y_{cl}(t') = \ddot{y}_{cl}(t') x_{cl}(t') + (2 \hbar/ m) x_{cl}(t') \int_{t'}^t a_I (t''-t') \, y_{cl}(t'') \, dt''$, which simplifies the classical action in \eqref{ClassAction1},
\be \label{ClassAction2}
	S_{cl} =
	 m \biggl[ x_f \dot{y}_{cl}(t) - x_i \dot{y}_{cl}(0)
	+ i \frac{\hbar}{m} \! \int \limits_0^t y_{cl}(t') \int \limits_0^t a_R(t'-t'') \, y_{cl}(t'') \, dt'' dt'
	\biggr] . 
\ee

Together, equations \eqref{ycl}, \eqref{denmatxy3}, \eqref{prefac} and \eqref{ClassAction2} give a general path integral formulation for the time evolution of a single quantum oscillator coupled to a thermal environment.  In this description, the thermal reservoir is modeled as a continuous spectrum of oscillators occupying a Boltzmann distribution at temperature $T$.  However, further progress requires an explicit form of $a_I (t''-t')$ which depends on the spectral density $J(\omega)$.
\section{Specifying the Spectral Density}

Typically, the spectrum of the reservoir is taken to have a linear form, $J(\omega) \propto \omega$, at low frequencies.  Since this spectrum cannot persist to arbitrarily high frequencies, there must be a cutoff beyond which there is no appreciable coupling to the quantum oscillator, $q$.  Because the thermal environment does not excite oscillator frequencies too far in excess of the mean thermal frequency, $(\beta \hbar)^{-1}$, it is reasonable to consider the case where the cutoff is twice this mean,
\be \label{J1}
	J(\omega) = 
	\begin{cases} 2 m \gamma \omega &  |\omega| < 2 /\beta \hbar  \\[.1in]
		0 & \text{otherwise}.
		\end{cases}
\ee
Whereas $2 \gamma$ is the usual damping parameter, the inverse of the mean thermal frequency, $\beta \hbar$, is identified as the thermalization time scale of the reservoir.  

Using the  spectral density of \eqref{J1} in the definition \eqref{aIdef1}, the expression for $a_I(t'-t'')$ can be written
\be \label{aI2}
	a_I(t'-t'') = -\frac{m \gamma}{ \pi i \hbar} \int \limits_{-2/\beta \hbar}^{2/\beta \hbar} \omega
	\, e^{i \omega (t'-t'')} \, d\omega .
\ee
Because the environmental reservoir contains many more degrees of freedom than the quantum oscillator, it is assumed that the thermalization time of the reservoir is always much shorter than $t$, the thermalization time of the oscillator.  Hence it is always the case that $2 t/ \beta \hbar \gg 1$ and the integration limits in \eqref{aI2} may be extended to infinity,
\begin{align} 
	a_I(t'-t'') &\simeq -\frac{m \gamma}{\pi i \hbar} \, 
	\int \limits_{-\infty}^{\infty} \omega
	\, e^{i \omega (t'-t'')} \, d\omega \nonumber \\
	&= \frac{2 m \gamma}{\hbar} \frac{d}{d t'} \, \delta(t'-t'') \label{aIdta} . 
\end{align}
Care must be exercised when taking this limit, for it implicitly contains the approximation
\be \label{a_approx}
	\frac{1}{2 \pi} \int \limits_{-2/\beta \hbar}^{2/ \beta \hbar} \! \! e^{i \omega (t'-t'')} \, d \omega
	\simeq \delta(t'-t'') \, .
\ee
When appearing within an integral, this approximation permits the delta function to be used.  However, when seemingly divergent expressions like $\delta(0)$ arise outside of any integral, it must be realized that what is really meant is
\be \label{delta0}
	\delta(0) = \frac{2}{\pi \beta \hbar} \, .
\ee

Simply substituting \eqref{aIdta} into the integro-differential equation \eqref{ycl} gives an ambiguous result since the delta function zero occurs at one of the integration limits.  Therefore, it is realized that the proper limit is obtained by half of the value of the corresponding case in which the delta function zero occurs within the integration limits.  Along with the observation in \eqref{delta0}, this prescription results in a second-order differential equation for $y_{cl}(t')$,
\be \label{ycl2}
	\ddot{y}_{cl}(t') - 2 \gamma \dot{y}_{cl}(t') + \left( \omega_0^{\,2} 
	- \frac{4 \gamma}{\pi \beta \hbar} \right) \! y_{cl}(t') = 0. 
\ee

\begin{figure} [h]
\begin{center}
\epsfxsize=3.5in\epsfbox{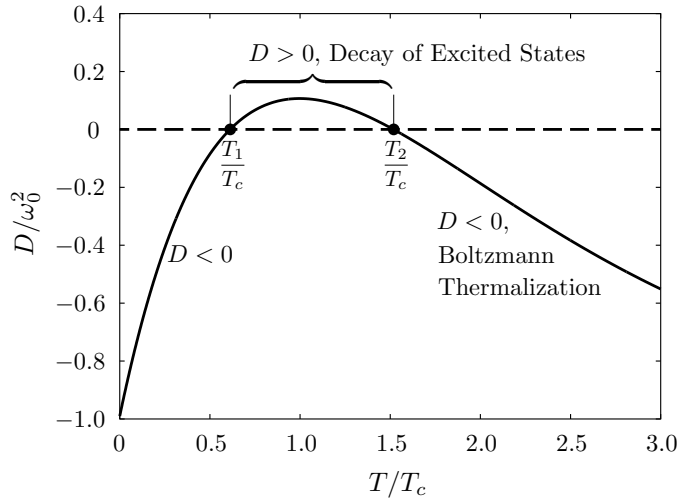}
\vspace{-.2in}
\end{center}
\caption {Using the damping parameter model of \eqref{epsilonmodel}, the associated discriminant \eqref{discrim} is divided by $\omega_0^2$, then plotted as a function of the normalized temperature, $T/T_c$, for $4 \gamma_0 k_B T_c/ (\pi e \hbar \, \omega_0^2) = 1.1$.  Although the quantum oscillator thermalizes to the Boltzmann distribution at low temperatures, there is a range $T_1 < T <T_2$ for which the excited oscillator states are suppressed since they decay into the stable ground state.  At temperatures beyond this range, the usual Boltzmann distribution re-emerges. \label{DPlot}}
\end{figure}

Whether the solutions to \eqref{ycl2} are exponential or oscillatory is determined by the discriminant 
\be \label{discrim}
	D = - \omega_0^{\,2} + \frac{4 \gamma}{\pi \beta \hbar} + \gamma^2 .
\ee
For the case of extreme underdamping, $\gamma \ll \omega_0$, the discriminant is close to $-\omega_0^2$ at low temperatures, thus giving oscillatory solutions.  However, if the temperature is sufficiently high that
\be \label{CoherenceCond}
	\frac{4 \gamma_0 k_B T}{\pi e \hbar \, \omega_0^2} > 1,
\ee
then the discriminant becomes positive, leading to purely exponential solutions of \eqref{ycl2}.  In the infinite-temperature limit, the term involving $\gamma T$ can only be convergent if $\gamma$ has a temperature dependence which decays at least as rapidly as $1/T$.  It is plausible to include these limits by taking an exponential form for the damping parameter
\be \label{epsilonmodel}
	\gamma = \gamma_0 e^{-T/T_c}, \qquad \gamma_0 \ll \omega_0 ,
\ee
in which $T_c$ is a temperature cutoff.  If the cutoff is sufficiently large then the discriminant \eqref{discrim} can be positive, thus giving purely exponential solutions around some neighborhood of $T \simeq T_c$.  As indicated by Fig.\,\ref{DPlot}, this solution gives rise to a unique steady-state, to be shown through corresponding derivations of the density matrix.
\section{Damping of Excited Modes}
Within some intermediate temperature range, the discriminant in \eqref{discrim} may be positive, indicating that the general solution to the differential equation \eqref{ycl2} has the form 
\be 
	y_{cl}(t') = e^{\gamma t'} \! \left\{ B \sinh(\xi t') + C \sinh[ \xi (t-t') ] \right\} ,
\ee
where $B$ and $C$ are constants to be determined from the boundary conditions and where we have defined
\be \label{xidef}
	\xi =  \sqrt{-\omega_0^2 + \frac{4 \gamma}{\pi \beta \hbar}  
	+ \gamma^2} > 0.
\ee
Applying the boundary conditions 
\begin{subequations} \label{BCs}
	\begin{align}
	y_{cl}(0) &= \frac{1}{\sqrt{2}} (q_i-q'_i) = y_i \\
	y_{cl}(t) &= \frac{1}{\sqrt{2}} (q_f - q'_f) = y_f , 
	\end{align}
\end{subequations}
gives
\be \label{yclsolnxi}
	y_{cl}(t') = \frac{1}{\sinh (\xi t)} \! \left\{ y_f e^{\gamma (t'-t)} \sinh(\xi t') 
	+ y_i e^{\gamma t'} \! \sinh[\xi (t-t')] \right\} .
\ee
The required end-point derivatives are obtained from \eqref{yclsolnxi},
\begin{subequations} \label{ydotxi0t}
	\begin{align}
	\dot{y}_{cl}(0) &= \gamma y_i + \frac{\xi}{\sinh(\xi t)} 
	\left[ y_f e^{-\gamma t} - y_i \cosh(\xi t) \right] \label{ydotxi0} \\
	\dot{y}_{cl}(t) &= \gamma y_f + \frac{\xi}{\sinh (\xi t)} 
	\left[ y_f\cosh(\xi t) - y_i e^{\gamma t} \right] . \label{ydotxit}
	\end{align}
\end{subequations}
In addition to the classical trajectory, $y_{cl}(t')$, the action is explicitly dependent upon $a_R(t'-t'')$, which is arrived at by inserting the spectral density \eqref{J1} into the definition \eqref{aRdef1},
\be \label{aR1}
	a_R (t'-t'') = \frac{\gamma m}{\pi \hbar} \! \int \limits_{-2/\beta \hbar}^{2/\beta \hbar} 
	\! \coth \! \left( \frac{\beta \hbar \omega}{2} \right) \omega \, e^{i \omega (t'-t'')} d \omega \, .
\ee
Expression \eqref{aR1} is now inserted into the corresponding integrals in the classical action \eqref{ClassAction2},
\begin{align} \label{aRInt}
	\frac{\hbar}{m} \int \limits_0^t y_{cl}(t') \int \limits_0^t a_R(t'-t'') \, y_{cl}(t'') \, dt'' dt' &= 
	\frac{\gamma}{\pi} \! \int \limits_{-2/\beta \hbar}^{2/\beta \hbar} \! d \omega
	\, \omega \coth \! \left( \frac{\beta \hbar \omega}{2} \right) \nonumber \\
	& \quad 
	\cdot \int \limits_0^t dt' e^{i \omega t'} y_{cl}(t') \int \limits_0^t dt'' e^{-i \omega t''} y_{cl}(t'') \, .
\end{align}
From the solution for $y_{cl}(t'')$ given by \eqref{yclsolnxi}, the required time integrals are evaluated to be
\begin{align} \label{I1xia}
	\int \limits_0^t dt'' e^{-i \omega t''} y_{cl}(t'') &= \frac{1}{2i} \frac{1}{\sinh (\xi t)} 
	\Biggl( y_f \! \left\{ \frac{1}{\omega - i(\xi - \gamma)} \left[ e^{-i (\omega - i\xi) t}
	- e^{-\gamma t} \right] \right.  \nonumber \\
	& \qquad \qquad \qquad \quad
	\left. \left. - \frac{1}{\omega + i(\xi + \gamma)} \left[ e^{-i (\omega + i\xi) t} 
	- e^{-\gamma t} \right] \right\} \right.  \nonumber \\
	& \qquad \qquad 
	- y_i \! \left\{ \frac{1}{\omega - i(\xi - \gamma)} \left[ e^{-i (\omega + i \gamma ) t}
	- e^{\xi t} \right] \right. \nonumber \\
	& \qquad \qquad \qquad 
	\left. -\frac{1}{\omega + i(\xi + \gamma)} \left[ e^{-i (\omega + i \gamma ) t}
	- e^{-\xi t} \right]  \right\} \Biggr) .
\end{align}
After multiplying this result by its complex conjugate, only the $\omega$ integral in \eqref{aRInt} remains, 
\begin{align} \label{aRIntSimpxi1}
	&\frac{\hbar}{m} \int \limits_0^t y_{cl}(t') \int \limits_0^t a_R(t'-t'') \, y_{cl}(t'') \, dt'' dt' \nonumber \\ 
	&= -\frac{1}{2} \frac{1}{\sinh^2(\xi t)} \frac{\gamma}{\pi} \! 
	\int \limits_{-2/\beta \hbar}^{2/\beta \hbar} \! d \omega
	\, \omega \coth \! \left( \frac{\beta \hbar \omega}{2} \right)  \! \bigl\{ y_f^2 e^{-\gamma t}
	\left[ f_\xi (\omega) - e^{-\xi t} g_\xi(\omega) \right] 
	\nonumber \\
	& \qquad \qquad \qquad \qquad \qquad \qquad \qquad \qquad
	-2 y_f y_i \left[ \cosh (\xi t) f_\xi(\omega) - g_\xi(\omega) \right] \nonumber \\[.1in]
	& \qquad \qquad \qquad \qquad \qquad \qquad \qquad \qquad
	+ y_i^2 e^{\gamma t} \left[ f_\xi(\omega) - e^{\xi t} g_\xi (\omega) \right] \bigr\}
	\nonumber \\[.1in]
	& \qquad \qquad \qquad \qquad \qquad \qquad \qquad \qquad \qquad
	+ \{ \xi \rightarrow - \xi \} .
\end{align}
For notational conciseness, the last term in curly braces, $\{ \xi \rightarrow - \xi \}$, is shorthand for the addition of the preceding expression but with $\xi$ replaced by $-\xi$.  Furthermore, the functions $f_\xi(\omega)$ and $g_\xi(\omega)$ are identified as
\begin{subequations} \label{fgDefs}
     \begin{align}
          f_\xi(\omega) &= \frac{1}{(\omega - i\xi)^2 + \gamma^2} \bigl\{ \cosh(\gamma t)
          - \cos \left[ (\omega - i\xi ) t \right] \bigr\} \label{fDef} \\
          g_\xi(\omega) &= \frac{1}{\omega^2 + (\xi - \gamma)^2} 
          \bigl\{ \cosh[(\xi - \gamma)t] - \cos (\omega t) \bigr\} . \label{gDef}
     \end{align}
\end{subequations}
In this regime, the required integrals over $\omega$ have been carried out in the appendix, where it is also shown that the thermalization limit $\gamma t \rightarrow \infty$ simplifies \eqref{aRIntSimpxi1} to
\be \label{aRIntSimpxi2}
	\frac{\hbar}{m} \int \limits_0^t y_{cl}(t') \int \limits_0^t a_R(t'-t'') \, y_{cl}(t'') \, dt'' dt'
	\xrightarrow[\gamma t \rightarrow \infty]{} i \gamma 
	\coth \! \left(\frac{i \beta \hbar \xi}{2} \right) \! \left( y_f^2 
	+ y_i^2 \right) .
\ee
In addition to \eqref{aRIntSimpxi2}, the corresponding long-time limits of \eqref{ydotxi0} and \eqref{ydotxit} are used in the classical action \eqref{ClassAction2} to obtain
\begin{align} \label{ClassAction3}
	S_{cl} &\xrightarrow[\gamma t \rightarrow \infty]{}
	\, m \! \left[ x_f (\gamma + \xi) y_f - x_i (\gamma - \xi) y_i 
	-\gamma \coth \! \left(\frac{i \beta \hbar \xi}{2} \right) \! 
	\left( y_f^2 + y_i^2 \right) \right]
	\nonumber \\
	& \qquad
	\simeq m \xi \! \left( x_f y_f + x_i y_i \right) ,
\end{align}
where the last line follows from the condition of severe underdamping in which $\gamma$ is by far the smallest frequency in the system.  Consequently, the entire contribution from \eqref{aRIntSimpxi2} is negligible.

Having obtained an expression for the effective action in terms of the end-point coordinates, we are now in a position to evaluate the integral $\int e^{(i/\hbar) S_{cl}}$ $\! \rho_0(x_i, y_i) dx_i dy_i$ occurring in the time-dependent density matrix \eqref{denmatxy3}.  The initial density matrix given by Eq.\,\eqref{denmat0} must be expressed in terms of the rotated coordinates \eqref{xydef},
\begin{align} \label{denmat0xy}
	\rho_0 (x_i, y_i) &= N_0 \exp \biggl\{ -\frac{m \omega_0}{2 \hbar} \biggl[ x_i^2 
	 \tanh \left( \frac{\beta_0 \hbar \omega_0}{2} \right) 
	 + y_i^2 \coth \left( \frac{\beta_0 \hbar \omega_0}{2} \right) \biggr] \biggr\} \, ,
\end{align} 
where the prefactor $N_0$ is identified as
\be \label{N01}
	N_0 = \sqrt{\frac{m \omega_0}{2 \pi \hbar \sinh( \beta_0 \hbar \omega_0)}} .
\ee
Using \eqref{ClassAction3} and \eqref{denmat0xy}, the steady state density matrix for intermediate temperatures can be written
\begin{align} \label{ssdenmatxi1}
	&\rho(x_f,y_f,t) \xrightarrow[\gamma t \rightarrow \infty]{} \nonumber \\
	&\frac{N_0}{Z_0} \, \mathfrak{T}(\gamma t \rightarrow \infty) 
	\int \limits_{-\infty}^{\infty} \int \limits_{-\infty}^{\infty} \exp \biggl\{ -\frac{m}{\hbar}
	\biggl[ \frac{1}{2} \, x_i^2 \, \omega_0 \tanh \! \left(\frac{\beta_0 \hbar \omega_0}{2} \right) 
	\nonumber \\
	& \qquad \qquad \qquad \qquad \qquad
	+ \frac{1}{2} \, y_i^2 \, \omega_0 \coth \! \left(\frac{\beta_0 \hbar \omega_0}{2} \right) 
	-i \xi x_i y_i  - i \xi x_f y_f \biggr] \biggr\} dx_i dy_i .
\end{align}
Evaluating the Gaussian integrals over $x_i$ and $y_i$, then using the definitions of $Z_0$ \eqref{Z02} and $N_0$ \eqref{N01} simplifies this to
\begin{align} \label{ssdenmatxi2}
	\rho(x_f,y_f,t) &\xrightarrow[\gamma t \rightarrow \infty]{} 
	2 \mathfrak{T}(\gamma t \rightarrow \infty) 
	\sqrt{\frac{\pi \hbar}{m \omega_0} \tanh \! \left(\frac{\beta_0 \hbar \omega_0}{2}\right)}
	\frac{\omega_0}{\sqrt{\omega_0^2 + \xi^2}} \nonumber \\
	& \qquad \quad  
	\cdot \exp \left(\frac{i}{\hbar} m \xi x_f y_f \right) .
\end{align}
It is instructive to use \eqref{xydef} to rewrite \eqref{ssdenmatxi2} in terms of the original wavefunction coordinates, $q_f$ and $q_f'$,
\begin{align} \label{ssdenmatxi3}
	\rho(x_f,y_f,t) &\xrightarrow[\gamma t \rightarrow \infty]{} 
	2 \mathfrak{T}(\gamma t \rightarrow \infty) 
	\sqrt{\frac{\pi \hbar}{m \omega_0} \tanh \! \left(\frac{\beta_0 \hbar \omega_0}{2}\right)}
	\frac{\omega_0}{\sqrt{\omega_0^2 + \xi^2}} \nonumber \\
	& \qquad \quad  
	\cdot \exp \left[-\frac{m}{2 \hbar} \left(-i\xi q_f^2 + i \xi q_f^{\prime \, 2} \right) \right] \nonumber \\
	&= 2 \mathfrak{T}(\gamma t \rightarrow \infty) 
	\sqrt{\frac{\pi \hbar}{m \omega_0} \tanh \! \left(\frac{\beta_0 \hbar \omega_0}{2}\right)}
	\frac{\omega_0}{\sqrt{\omega_0^2 + \xi^2}} \sqrt{\frac{\pi \hbar}{m \xi}} 
	\, \varphi_0(q_f) \varphi_0^\ast (q_f') ,
\end{align}
where the wave function $\varphi_0(q_f)$ has been defined by 
\be
	\varphi_0(q_f) = \left(\frac{-i m \xi}{\pi \hbar} \right)^{\! 1/4} \!
	\exp\left( \frac{i m \xi}{2 \hbar} q_f^2 \right) .
\ee
\indent Therefore, in this higher temperature regime the equilibrium state of the harmonic oscillator is characterized by its ground state, but with the imaginary frequency $-i \xi$.  If terms of order $\gamma$ were not dropped, this frequency would be complex-valued in general.  Because the imaginary part of the frequency quantifies the decay rate into the stable ground state, all excited modes are unstable except for the ground state where the energy can always be chosen to be real \cite{Rajeev}.  Since this decay rate is linearly related to the energy quantum number $n$, the associated instability becomes more prominent with increasing excitation.  This result is interpreted as the reservoir acting to dissipate or extract energy from the oscillator, thereby preserving the coherence of a single quantum state.  
\section{Thermalization to a Boltzmann Distribution}
Through the decay of excited states, the environment can act to maintain the coherence of the oscillator at intermediate temperatures.  By contrast, at low or at very high temperatures the discriminant in \eqref{discrim} becomes negative, thus changing the solutions of \eqref{ycl2} from exponential to oscillatory,  
\be 
	y_{cl}(t') = e^{\gamma t'} \! \left\{ B \sin(\eta t') + C \sin[ \eta (t-t') ] \right\} ,
\ee
where the frequency is given by 
\be \label{etadef}
	\eta =  \sqrt{\omega_0^2 - \frac{4 \gamma}{\pi \beta \hbar}  
	- \gamma^2} > 0.
\ee
Analysis that parallels that of the intermediate temperature case need not be repeated.  Instead, corresponding expressions are obtained by the replacement $\xi \rightarrow -i \eta$ in the previous treatment.  Using this substitution in \eqref{yclsolnxi} and \eqref{ydotxi0t} determines the analogous solution for this case,
\be \label{yclsolneta} 
     	y_{cl}(t') = \frac{1}{\sin (\eta t)} \! \left\{ y_f e^{\gamma (t'-t)} \sin(\eta t') 
	+ y_i e^{\gamma t'} \! \sin[\eta (t-t')] \right\} ,
\ee
with the corresponding end-point velocities given by
\begin{subequations} 
     \begin{align}
	\dot{y}_{cl}(0) &\simeq  \frac{\eta}{\sin(\eta t)} 
	\left[ y_f e^{-\gamma t} - y_i \cos(\eta t) \right] \label{ydoteta0} \\
	\dot{y}_{cl}(t) &\simeq \frac{\eta}{\sin (\eta t)} 
	\left[ y_f\cos(\eta t) - y_i e^{\gamma t} \right] . \label{ydotetat}
	\end{align}
\end{subequations}
Arriving at \eqref{ydoteta0} and \eqref{ydotetat} follows from the fact that all calculations are performed in the severely underdamped regime for $\gamma \ll \xi, \eta$.   Additionally, the expression for the integral of $a_R(t'-t'')$ is obtained by first substituting \eqref{fgDefs} into \eqref{aRIntSimpxi1}, then again using the transformation $\xi \rightarrow -i \eta$,
\begin{align} \label{aRIntSimpeta1}
	&\frac{\hbar}{m} \int \limits_0^t y_{cl}(t') \int \limits_0^t a_R(t'-t'') \, y_{cl}(t'') \, dt'' dt' \nonumber \\ 
	&= \frac{\gamma}{\pi} \! \int \limits_{-2/\beta \hbar}^{2/\beta \hbar} \! d \omega
	\, \omega \coth \! \left( \frac{\beta \hbar \omega}{2} \right) \frac{1}{2} \frac{1}{\sin^2(\eta t)}
	\nonumber \\
	& \qquad \quad 
	\cdot \Biggl[ y_f^2 \Biggl( \frac{1}{(\omega - \eta)^2 + 
	\gamma^2} \, e^{-\gamma t} \! \left\{ \cosh(\gamma t) - \cos[ (\omega - \eta) t] 
	\right\} \nonumber \\
	& \qquad \qquad \quad
	- \frac{1}{\omega^2 - (\eta - i\gamma)^2} \, e^{(i\eta - \gamma)t} \!
	\left\{ \cos[(\eta - i\gamma)t] - \cos (\omega t) \right\} \Biggr) \nonumber \\[.1in]
	& \qquad 
	-2 y_f y_i \Biggl( \frac{1}{(\omega - \eta)^2 + \gamma^2} \cos(\eta t) \! \left\{
	\cosh(\gamma t) - \cos[(\omega - \eta)t] \right\} \nonumber \\
	& \qquad \qquad \quad 
	- \frac{1}{\omega^2  - (\eta - i\gamma)^2}
	\left\{ \cos[(\eta - i\gamma)t] - \cos(\omega t) \right\} \Biggr) \nonumber \\[.1in]
	& \qquad \quad 
	+ y_i^2 \Biggl( \frac{1}{(\omega - \eta)^2 + \gamma^2} \, e^{\gamma t} \!
	\left\{\cosh(\gamma t) - \cos[ (\omega - \eta) t] \right\} \nonumber \\
	& \qquad \qquad \quad
	- \frac{1}{\omega^2 - (\eta - i\gamma)^2} \, e^{(-i\eta + \gamma)t} \!
	\left\{ \cos[(\eta - i\gamma)t] - \cos (\omega t) \right\}  \Biggr) \Biggr]\nonumber \\
	& 	
	+ \{ \eta \rightarrow - \eta \} .
\end{align}
Due to the oscillatory nature of the solution for $y_{cl}(t')$, the long-time approximations that led to \eqref{aRIntSimpxi2} do not apply for this case.  Nonetheless, the damping parameter remains the smallest frequency in the system, thereby allowing terms of order $\gamma$ and higher to be separated out as negligible.  Hence, the dominant terms in the integrals over $\omega$ are easily obtained
\begin{subequations} \label{I1I2}
\begin{align} 
	& \frac{\gamma}{\pi} \! \int \limits_{-2/\beta \hbar}^{2/\beta \hbar} \! d \omega \, \omega 
	\coth \! \left( \frac{\beta \hbar \omega}{2} \right) \!
	\frac{1}{(\omega - \eta)^2 + \gamma^2} \cos[ (\omega - \eta) t]  \nonumber \\
	&= \frac{\gamma}{\pi} 
	Re \left\{ \int \limits_{-2/\beta \hbar}^{2/\beta \hbar} \! d \omega \, \omega 
	\coth \! \left( \frac{\beta \hbar \omega}{2} \right) \!
	\frac{1}{(\omega - \eta - i \gamma)(\omega - \eta + i \gamma)}
	e^{i (\omega - \eta) t} \right\} \nonumber \\
	&= \eta \coth \! \left( \frac{\beta \hbar \eta}{2} \right) \! e^{-\gamma t} 
	+ \{ \text{terms of order} \geq \gamma \} , \label{I1} \\[.1in]
	& \frac{\gamma}{\pi} \! \int \limits_{-2/\beta \hbar}^{2/\beta \hbar} \! d \omega \, \omega 
	\coth \! \left( \frac{\beta \hbar \omega}{2} \right) \!
	\frac{1}{\omega^2 - (\eta - i\gamma)^2} \cos(\omega t)  \nonumber \\
	&=  \{ \text{terms of order} \geq \gamma \} . \label{I2}
\end{align}
\end{subequations}
All remaining integrals appearing in \eqref{aRIntSimpeta1} are found from the $t=0$ forms of \eqref{I1} and \eqref{I2}.
Using these results, the required integral involving $a_R(t'-t'')$ becomes
\begin{align} \label{aRIntSimpeta2}
	&\frac{\hbar}{m} \int \limits_0^t y_{cl}(t') \int \limits_0^t a_R(t'-t'') \, y_{cl}(t'') \, dt'' dt' 
	\nonumber \\ 
	&
	= \frac{\eta}{\sin^2(\eta t)} \coth \! \left( \frac{\beta \hbar \eta}{2} \right) \sinh(\gamma t)
	\! \left[ y_f^2 e^{-\gamma t} - 2 y_i y_f \cos(\eta t) + y_i^2 e^{\gamma t}
	\right].
\end{align}
Substituting \eqref{ydoteta0}, \eqref{ydotetat} and \eqref{aRIntSimpeta2} into the expression for the classical action \eqref{ClassAction2} leads to the result
\begin{align} \label{ClassActioneta1}
	S_{cl} &= m \frac{\eta}{\sin (\eta t)} \biggl\{ x_f y_f \cos (\eta t) 
	- x_f y_i \, e^{\gamma t} - x_i y_f e^{-\gamma t} + x_i y_i \cos(\eta t) 
	\nonumber \\
	&
	+\frac{i}{\sin(\eta t)} \coth \! \left( \frac{\beta \hbar \eta}
	{2} \right) \sinh(\gamma t) \! \left[ y_f^2 e^{- \gamma t} - 2y_i y_f \cos(\eta t) 
	+ y_i^2 e^{\gamma t} \right] \biggr\} . \nonumber \\[-.08in]
\end{align}
As in the previous analysis, obtaining the effective action in terms of the end-point coordinates enables the evaluation of the integral $\int e^{(i/\hbar) S_{cl}} \rho_0(x_i, y_i) dx_i dy_i$ appearing in the time-dependent density matrix \eqref{denmatxy3}.  Inserting \eqref{ClassActioneta1} and \eqref{denmat0xy} into the $x_i$ and $ y_i$ integrals gives
\begin{align} \label{denmatfactor1}
	&\int e^{\frac{i}{\hbar} S_{cl}} \rho_0(x_i,y_i) dx_i dy_i \nonumber \\
	&= N_0 \int \limits_{-\infty}^\infty \int \limits_{-\infty}^\infty
	\exp \biggl\{ -\frac{m}{\hbar}
	\biggl[ \frac{1}{2} \, x_i^2 \, \omega_0 \tanh \! \left(\frac{\beta_0 \hbar \omega_0}{2} \right) 
	+ \frac{1}{2} \, y_i^2 \, \omega_0 \coth \! \left(\frac{\beta_0 \hbar \omega_0}{2} \right) 
	\nonumber \\
	& \qquad \qquad \qquad \qquad
	-i x_f y_f \eta \cot( \eta t) + i x_f y_i \eta \csc(\eta t) 
	e^{\gamma t} \nonumber \\
	& \qquad \qquad \qquad \qquad
	+ ix_i y_f \eta \csc( \eta t) e^{-\gamma t} 
	-ix_i y_i \eta \cot(\eta t) \nonumber \\
	& \qquad \qquad \qquad \qquad
	+  y_f^2 \, \eta \csc^2(\eta t) \coth \! 
	\left( \frac{\beta \hbar \eta}{2} \right) 
	\sinh( \gamma t) e^{-\gamma t} \nonumber \\
	& \qquad \qquad \qquad  \qquad
	- 2 y_i y_f \eta \cot(\eta t) \csc(\eta t) 
	\coth \! \left(\frac{\beta \hbar \eta}{2}\right) \sinh(\gamma t) \nonumber \\
	& \qquad \qquad \qquad \qquad
	+ y_i^2 \, \eta \csc^2(\eta t) 
	\coth\! \left(\frac{\beta \hbar \eta}{2}\right) \sinh(\gamma t) e^{\gamma t} 
	\biggr] \biggr\} dx_i dy_i .
\end{align}
Performing the Gaussian integrals along with the lengthy algebra involved leaves a simplified result
\begin{align} \label{denmatfactor2}
	&\int e^{\frac{i}{\hbar} S_{cl}} \rho_0(x_i,y_i) dx_i dy_i  \nonumber \\
	&= N_0 \frac{2 \pi \hbar}{m} \sqrt{\frac{z_0}{\omega_0 \nu}} 
	\exp \Biggl( -\frac{m}{2 \hbar \nu} \biggl\{ y_f^2 \, \eta^2 \csc^2(\eta t)
	\biggl[  4 z^2 \sinh^2 (\gamma t) \nonumber \\
	& \qquad \qquad \qquad \qquad \qquad
	+2 \! \left( \frac{\eta}{\omega_0} + \frac{\omega_0}{\eta} \right) \! z
	z_0 \sinh(\gamma t) e^{-\gamma t} + z_0^2 \, e^{-2 \gamma t} \biggr] \nonumber \\
	& \qquad \qquad \qquad
	+ 2i x_f y_f \biggl(\frac{\eta^2}{\omega_0} - \omega_0 \biggr) \eta \, z_0
	\cot(\eta t) + x_f^2 \eta^2 \csc^2 (\eta t) e^{2 \gamma t} \biggr\} \Biggr) ,
\end{align}
in which the parameters
\begin{subequations}
\begin{align}
	z &= \coth\! \left(\frac{\beta \hbar \eta}{2} \right) \label{zdef} \\
	z_0 &= \coth\! \left(\frac{\beta_0 \hbar \omega_0}{2} \right) \label{z0def} 
\end{align}
\end{subequations}
and
\be \label{nudef}
	\nu = \omega_0 z_0
	+ \frac{\eta^2}{\omega_0} z_0 \cot^2(\eta t)  
	+ 2 \eta z \csc^2 (\eta t) \sinh(\gamma t) e^{\gamma t} , 
\ee
have been defined for brevity.
Assuming $\gamma$ to be sufficiently small, the renormalized frequency in \eqref{etadef} can be approximated by
\be \label{eta=w0}
	\eta \simeq \omega_0.
\ee
Being in the highly underdamped regime, the expression for $\nu$ simplifies to
\be \label{nusimp}
	\nu \simeq \omega_0 \csc^2(\omega_0 t) 
	\biggl[ \coth \! \left(\frac{\beta_0 \hbar \omega_0}{2} \right) 
	+ 2 \coth \! \left(\frac{\beta \hbar \omega_0}{2} \right)  
	\! \sinh(\gamma t) e^{\gamma t} \biggr] ,
\ee
where we have inserted the explicit expressions for $z$ and $z_0$.  Likewise, \eqref{eta=w0} also simplifies expression \eqref{denmatfactor2} 
\begin{align} \label{denmatfactor3}
	&\int e^{\frac{i}{\hbar} S_{cl}} \rho_0(x_i,y_i) dx_i dy_i \nonumber \\
	&= N_0 \frac{2 \pi \hbar}{m} 
	\sqrt{\frac{z_0}{\omega_0 \nu}} \exp \Biggl\{ -\frac{m \omega_0}{2 \hbar} \biggl[ 
	x_f^2 \frac{\omega_0}{\nu \sin^2 (\omega_0 t)} e^{2 \gamma t} 
	+ y_f^2 \, \frac{\nu}{\omega_0} 
	\sin^2(\omega_0 t) e^{-2 \gamma t} \biggr] \Biggr\} . 
\end{align}
Substituting this result into \eqref{denmatxy3} gives the time evolution of the full density matrix
\begin{align} \label{denmat3}
	\rho(x_f,y_f,t) = 2 \mathfrak{T}(t) \sqrt{\frac{\pi \hbar}{m \nu}}
	\exp \Biggl\{ -\frac{m \omega_0}{2 \hbar} \biggl[ 
	&x_f^2 \frac{\omega_0}{\nu \sin^2 (\omega_0 t)} e^{2 \gamma t} \nonumber \\
	& \quad 
	+ y_f^2 \, \frac{\nu}{\omega_0} 
	\sin^2(\omega_0 t) e^{-2 \gamma t} \biggr] \Biggr\} .
\end{align}
Since no long-time limits have been used in the derivation, equation \eqref{denmat3} includes the transient time behavior of the density matrix.  In addition to the expression for $\nu$ \eqref{nusimp}, this result shows the thermalization time to be on the order of the inverse damping parameter, $t \sim \gamma^{-1}$, just like the previous case for intermediate temperatures.  At times in excess of $\gamma^{-1}$, $\nu$ simplifies to
\be \label{nulimit}
	\nu \xrightarrow[\gamma t \rightarrow \infty]{} \omega_0 \csc^2(\omega_0 t) 
	\coth \! \left( \frac{\beta \hbar \omega_0}{2} \right) \! e^{2 \gamma t} .
\ee
This limit demonstrates that when in contact with the thermalizing environment, the quantum oscillator equilibriates to a steady-state described by the density matrix
\begin{align} \label{denmat4}
	\rho(x_f,y_f,\gamma t \rightarrow \infty) &= 2 \mathfrak{T}(\gamma t \rightarrow \infty) 
	\sqrt{\frac{\pi \hbar}{m \omega_0}\tanh \! \left(\frac{\beta \hbar \omega_0}{2} \right)} \, 
	\sin (\omega_0 t) \, e^{-\gamma t}
	\nonumber \\
	& \qquad
	\cdot \exp \Biggl\{ -\frac{m \omega_0}{2 \hbar} \biggl[ 
	x_f^2 \tanh \! \left(\frac{\beta \hbar \omega_0}{2} \right) 
	+ y_f^2 \coth \! \left(\frac{\beta \hbar \omega_0}{2} \right) \biggr] \Biggr\} .
	\nonumber \\[-.08in]
\end{align}
Comparison with the initial density matrix of \eqref{denmat0xy} indicates that at low and at very high temperatures, for which the discriminant in \eqref{discrim} is negative, the reservoir acts to thermalize the quantum oscillator to the usual Boltzmann distribution of eigenstates.  
\section{Summary}
The steady-state of a quantum oscillator can be dependent on the temperature of the environment to which it is coupled.  Modeled as an oscillator continuum, the environment is quantified by a spectral density which is linear in frequency with a proportionality constant known as the coupling, or damping, parameter.  Because it is improbable to excite energies too far in excess of the mean thermal energy, the environmental spectrum has a cutoff of twice the mean thermal frequency, $2/\beta \hbar$.  

In the path integral formulation, the steady-state of the oscillator is dependent upon the renormalized frequency, $\sqrt{\omega_0^2 - 4 \gamma / (\pi \beta \hbar) -\gamma^2}$, which inherits its temperature dependence from the environmental cutoff.  For low temperatures, the small frequency shift, $\simeq -2 \gamma / (\pi \beta \hbar \omega_0) -\gamma^2/(2 \omega_0)$, obtains the expected Boltzmann thermalization.  To keep the renormalized frequency finite in the high-temperature limit, a damping parameter ($2 \gamma$) with a temperature cutoff, $T_c$, is considered.  If this cutoff is small enough such that the renormalized frequency is always real, then the expected Boltzmann thermalization prevails at any temperature.  However, if $T_c$ is sufficiently large, there is a range of temperatures for which the renormalized frequency becomes imaginary.  Decaying at a rate given by the corresponding imaginary energy eigenvalue, all excited modes become unstable, whereas the ground state remains stable since its energy can always be chosen to be real.  Dissipation now refers to an extraction of energy instead of a suppression of coherence.  

Concisely summarized by the example in Fig.\,\ref{DPlot}, the usual Boltzmann thermalization prevails at low and at very high temperatures.  For temperatures in between the environment damps out the excited modes, thereby maintaining the coherence of a single quantum state.  As indicated by experiments on photosynthetic molecules \cite{EngelFleming, PanitEngel} and by recent experiments on superconducting graphite grains \cite{Esquinazi}, there are systems where this intermediate damping regime could lie well within the room-temperature range.  In particular, photosynthetic molecules display a quantum coherence having the same qualitative temperature dependence as described here.  Such dependencies also arise in models of the exciton transport efficiency \cite{Lloyd, Pleino}, thus suggesting a possible mechanism for a high-temperature cutoff in the damping parameter.  

\appendix
\section{Frequency Integrals for the Intermediate Temperature Region} 
Due to the appearance of hyperbolic functions, obtaining the dominant terms in the $\omega$ integrals of \eqref{aRIntSimpxi1} is somewhat more involved than in the very-high or in the lower temperature counterparts.  It is convenient to first consider the integral involving $f_\xi(\omega)$ which contains two terms of the form
\begin{align} \label{A1}
	&\frac{\gamma}{\pi} \! \int \limits_{-2/\beta \hbar}^{2/\beta \hbar} \! d \omega
	\, \omega \coth \! \left( \frac{\beta \hbar \omega}{2} \right) \frac{1}{(\omega - i\xi)^2 
	+ \gamma^2} \cos \left[ (\omega - i\xi)t \right] \nonumber \\
	\begin{split}
	&= \frac{1}{2 \pi i} \! \int \limits_{-2/\beta \hbar}^{2/\beta \hbar} \! d \omega
	\, \omega \coth \! \left( \frac{\beta \hbar \omega}{2} \right) \!
	\left[\frac{1}{\omega - i(\xi + \gamma)}
	- \frac{1}{\omega - i(\xi - \gamma)} \right] \\
	&\qquad \qquad \qquad \qquad \qquad \qquad
	\cdot \left[ \cosh(\xi t) \cos(\omega t) + i \sinh(\xi t) \sin(\omega t) \right] . 
	\end{split}
\end{align}
Since this integral splits into two terms of opposite parity in $\gamma$, it suffices to evaluate only one, with the other obtained from the substitution $\gamma \rightarrow - \gamma$.  A more compact expression can be written by denoting $(i)$ as the integral multiplying $\cosh(\xi t)$ and $(ii)$ as the integral multiplying $\sinh(\xi t)$,
\begin{align} \label{A2}
	&\int \limits_{-2/\beta \hbar}^{2/\beta \hbar} \! d \omega
	\, \omega \coth \! \left( \frac{\beta \hbar \omega}{2} \right) \!
	\frac{1}{\omega - i(\xi + \gamma)} 
	\left[ \cosh(\xi t) \cos(\omega t) + i \sinh(\xi t) \sin(\omega t) \right] \nonumber \\
	& \qquad \qquad
	= \cosh(\xi t) (i) + i \sinh(\xi t) (ii) .
\end{align}
Reading off the integral identified as $(i)$ we have
\begin{align} \label{A3}
	(i) &= \int \limits_{-2/\beta \hbar}^{2/\beta \hbar} \! d \omega
	\, \omega \coth \! \left( \frac{\beta \hbar \omega}{2} \right) \!
	\frac{1}{\omega - i(\xi + \gamma)} 
	\cos(\omega t) \nonumber \\
	&= i(\xi + \gamma) \! \! \int \limits_{-2/\beta \hbar}^{2/\beta \hbar} \! d \omega
	\, \omega \coth \! \left( \frac{\beta \hbar \omega}{2} \right) \!
	\frac{1}{\left[\omega + i(\xi + \gamma)\right] \!
	\left[\omega - i(\xi + \gamma) \right]} 
	\cos(\omega t) \nonumber \\
	&= i(\xi + \gamma) Re \left\{ \int \limits_{-2/\beta \hbar}^{2/\beta \hbar} \! d \omega
	\, \omega \coth \! \left( \frac{\beta \hbar \omega}{2} \right) \!
	\frac{1}{\left[\omega + i(\xi + \gamma)\right] \!
	\left[\omega - i(\xi + \gamma) \right]} 
	\, e^{i \omega t} \right\} \nonumber \\
	&= i(\xi + \gamma) Re \left\{ \vphantom{\int \limits_{{\cal C}_T}}
	2 \pi i [i (\xi + \gamma)]
	\coth \! \left[ \frac{i \beta \hbar (\xi + \gamma)}{2} \right] \!
	\frac{1}{2i (\xi + \gamma)} 
	\, e^{-(\xi + \gamma) t} \right. \nonumber \\
	& \qquad \qquad \qquad \qquad 
	\left. - \int \limits_{{\cal C}_T} d \omega
	\, \omega \coth \! \left( \frac{\beta \hbar \omega}{2} \right) \!
	\frac{1}{\omega^2 + (\xi + \gamma)^2} \, e^{i \omega t} \right\} .
\end{align}
Arriving at the last line requires an integration over contour ${\cal C}_T$, the positively-oriented semicircular arc of radius $2/\beta \hbar$ in the upper half complex plane.  In the intermediate regime, temperatures are sufficiently high that
\be \label{A4}
	\gamma, \xi \ll (\beta \hbar)^{-1} ,
\ee
thus allowing the integral over ${\cal C}_T$ to be approximated as
\be \label{A5}
	\int \limits_{{\cal C}_T} d \omega
	\, \omega \coth \! \left( \frac{\beta \hbar \omega}{2} \right) \!
	\frac{1}{\omega^2 + (\xi + \gamma)^2} \, e^{i \omega t}
	\lesssim
	\int \limits_{{\cal C}_T} d \omega
	\, \frac{1}{\omega} \coth \! \left( \frac{\beta \hbar \omega}{2} \right) \! e^{i \omega t} .
\ee
The contour ${\cal C}_T$ is more conveniently parametrized by the change of variables
\begin{align} \label{A6}
	\omega &= r e^{i \theta} , \quad r = \frac{2}{\beta \hbar} \nonumber \\
	d \omega &= i \omega d \theta.
\end{align}
With these substitutions, the upper bound \eqref{A5} becomes	
\begin{align} \label{A7}
	\int \limits_{{\cal C}_T} d \omega
	\, \frac{1}{\omega} \coth \! \left( \frac{\beta \hbar \omega}{2} \right) \! e^{i \omega t} 
	&= i \int \limits_0^\delta d\theta \coth \! \left(e^{i \theta} \right) \! e^{i r e^{i \theta} t} 
	+ i \int \limits_\delta^{\pi - \delta} d \theta \coth \! \left(e^{i \theta} \right) \! e^{i r e^{i \theta} t}
	\nonumber \\
	& 
	+ i \int \limits_{\pi - \delta}^\pi d \theta \coth \! \left(e^{i \theta} \right) \! e^{i r e^{i \theta} t} ,
\end{align}
where the small parameter $\delta$ has been defined as
\be \label{A8}
	\delta = \beta \hbar (\xi + \gamma) \ll 1.
\ee
To leading order, of the sum of the first integral and the last integral is linear in $\delta$,
\begin{align} \label{A9}
	\int \limits_0^\delta d\theta \coth \! \left(e^{i \theta} \right) \! e^{i r e^{i \theta} t} 
	+  \int \limits_{\pi - \delta}^\pi d \theta \coth \! \left(e^{i \theta} \right) \! e^{i r e^{i \theta} t}
	& \simeq \coth(1) e^{i r t} \delta + \coth(-1) e^{-i r t} \delta \nonumber \\
	&= 2 i \coth(1) \sin \! \left(\frac{2t}{\beta \hbar} \right) \delta .
\end{align}
For the remaining term in \eqref{A7}, an upper bound can be established since
\be \label{A10}
	\int \limits_\delta^{\pi - \delta} d\theta \coth \! \left(e^{i \theta} \right) \! e^{i r e^{i \theta} t}
	< e^{-2(\xi + \gamma) t} \int \limits_\delta^{\pi - \delta} d\theta \coth \! \left(e^{i \theta} \right) 
	\! e^{i r \cos(\theta) t}.
\ee
Terms of order $e^{-\gamma t}$ are retained since these are used to obtain the corresponding terms of opposite parity in $\gamma$.  

Because all calculations are performed in the severely underdamped case, the damping parameter $\gamma$ is by far the smallest frequency in the system.  Consequently, the integral in \eqref{A10} may be neglected as it vanishes exponentially as $e^{-2 \xi t} \ll e^{-\gamma t}$ in the required large-time limit.  From these observations, the leading order contribution of the ${\it C}_T$ contour integral is
\be \label{A11}
	\int \limits_{{\cal C}_T} d \omega
	\, \omega \coth \! \left( \frac{\beta \hbar \omega}{2} \right) \!
	\frac{1}{\omega^2 + (\xi + \gamma)^2} \, e^{i \omega t}
	\simeq
	-2 \coth(1) \sin \! \left(\frac{2t}{\beta \hbar}\right) \delta .
\ee
Using \eqref{A11} in \eqref{A3}, the dominant terms of integral $(i)$ are
\begin{align} \label{A12}
	(i) &= \int \limits_{-2/\beta \hbar}^{2/\beta \hbar} \! d \omega
	\, \omega \coth \! \left( \frac{\beta \hbar \omega}{2} \right) \!
	\frac{1}{\omega - i(\xi + \gamma)} \cos(\omega t) \nonumber \\
	\begin{split}
	&= (\xi + \gamma) \Biggl\{-\pi \coth \! \left[\frac{i \beta \hbar (\xi + \gamma)}{2} \right]
	\! e^{-(\xi + \gamma)t} \\
	& \quad
	+ 2i \coth(1) \sin \! \left( \frac{2t}{\beta \hbar} \right) \!
	\beta \hbar (\xi + \gamma) + \mathcal{O} \! \left(\beta \hbar \xi \right)^2 \Biggr\}.
	\end{split}
\end{align}
The second integral of \eqref{A2} is identified as
\begin{align} \label{A13}
	(ii) &= \int \limits_{-2/\beta \hbar}^{2/\beta \hbar} \! d \omega
	\, \omega \coth \! \left( \frac{\beta \hbar \omega}{2} \right) \!
	\frac{1}{\omega - i(\xi + \gamma)} 
	\sin(\omega t) \nonumber \\
	&= \int \limits_{-2/\beta \hbar}^{2/\beta \hbar} \! d \omega
	\, \omega^2 \coth \! \left( \frac{\beta \hbar \omega}{2} \right) \!
	\frac{1}{\omega^2 + (\xi + \gamma)^2} 
	\sin(\omega t) \nonumber \\
	&= \frac{i}{\xi + \gamma} \frac{\partial}{\partial t} (i) \nonumber \\
	&= (\xi + \gamma) \Biggl\{\pi i \coth \! \left[\frac{i \beta \hbar (\xi + \gamma)}{2} \right] \!
	e^{-(\xi + \gamma)t} \nonumber \\
	& \quad
	- 4 \coth(1) \cos \! \left(\frac{2t}{\beta \hbar} \right) 
	+ \mathcal{O} \! \left(\beta \hbar \xi \right) \Biggr\}
\end{align}
Results \eqref{A12} and \eqref{A13} are used to express \eqref{A2} in terms of the required parameters
\begin{align} \label{A14}
	\cosh(\xi t) (i) + i \sinh(\xi t) (ii) 
	&= -(\xi + \gamma) \Biggl\{4i \coth(1) \cos \! \left(\frac{2 t}{\beta \hbar} \right) \! \sinh(\xi t) 
	\nonumber \\
	& \! \! \! +\pi \coth \! \left[\frac{i \beta \hbar (\xi + \gamma)}{2} \right] \!
	e^{-\gamma t} + \mathcal{O} \! \left[\beta \hbar \xi \sinh(\xi t)\right] \Biggr\}
\end{align}

As noted, the second contribution on the right-hand side of \eqref{A1} is derived by replacing $\gamma$ with $- \gamma$ in \eqref{A14}.  Subtracting these two contributions, then dividing by $2 \pi i$ obtains \eqref{A1} 
\begin{align} \label{A15}
	\frac{\gamma}{\pi} \! \int \limits_{-2/\beta \hbar}^{2/\beta \hbar} \! &d \omega
	\, \omega \coth \! \left( \frac{\beta \hbar \omega}{2} \right) \frac{1}{(\omega - i\xi)^2 
	+ \gamma^2} \cos \left[ (\omega - i\xi)t \right] \nonumber \\
	&
	= \frac{i}{2} (\xi + \gamma) \coth \! \left[\frac{i \beta \hbar (\xi + \gamma)}{2} \right] \!
	e^{-\gamma t} - \frac{i}{2} (\xi - \gamma) 
	\coth \! \left[\frac{i \beta \hbar (\xi - \gamma)}{2} \right] \! e^{\gamma t} \nonumber \\[.1in]
	& \qquad
	-\frac{4 \gamma}{\pi} \! \left[ \coth(1) \cos \! \left(\frac{2t}{\beta \hbar} \right) 
	+ \mathcal{O} \! \left( \beta \hbar \xi \right) \right] \! \sinh(\xi t) .
\end{align}
Equation \eqref{A15} along with its $t=0$ form are used to obtain the $f_\xi(\omega)$ integral in \eqref{aRIntSimpxi1},
\begin{align} \label{A16}
	\frac{\gamma}{\pi} \! \int \limits_{-2/\beta \hbar}^{2/\beta \hbar} \! &d \omega
	\, \omega \, f_\xi(\omega) \coth \! \left( \frac{\beta \hbar \omega}{2} \right) \nonumber \\
	&= \frac{i}{2} \Biggl\{ (\xi + \gamma) \coth \! \left[\frac{i \beta \hbar (\xi + \gamma)}{2} \right] \!
	+ (\xi - \gamma) \coth \! \left[\frac{i \beta \hbar (\xi - \gamma)}{2} \right] \Biggr\} \sinh(\gamma t)
	\nonumber \\
	& \qquad \qquad
	+ \frac{4 \gamma}{\pi} \left[ \coth(1) \cos\left(\frac{2t}{\beta \hbar} \right) 
	+ \mathcal{O} \! \left( \beta \hbar \xi \right) \right] \sinh(\xi t) .
\end{align}

Also in \eqref{aRIntSimpxi1} is the integral of $g_\xi(\omega)$ which is a linear combination of two terms of the form
\begin{align} \label{A17}
	\frac{\gamma}{\pi}& \! \int \limits_{-2/\beta \hbar}^{2/\beta \hbar} \! d \omega
	\, \omega \coth \! \left( \frac{\beta \hbar \omega}{2} \right) \frac{1}{\omega^2 
	+ (\xi - \gamma)^2} \cos (\omega t) \nonumber \\
	&
	= \frac{\gamma}{\pi} Re \left\{ \int \limits_{-2/\beta \hbar}^{2/\beta \hbar} \! d \omega
	\, \omega \coth \! \left( \frac{\beta \hbar \omega}{2} \right) \!
	\frac{1}{\left[\omega + i(\xi - \gamma) \right] \! \left[\omega - i(\xi - \gamma) \right]}
	\, e^{i \omega t} \right\} \nonumber \\
	&
	= \frac{\gamma}{\pi} Re \left\{ \vphantom{\int \limits_{{\cal C}_T}}
	\pi i \coth \! \left[ \frac{i \beta \hbar (\xi - \gamma)}{2} \right] \!
	e^{-(\xi - \gamma) t} \right. \nonumber \\
	& \qquad \qquad \qquad \qquad  \qquad
	\left. - \int \limits_{{\cal C}_T} d \omega
	\, \omega \coth \! \left( \frac{\beta \hbar \omega}{2} \right) \!
	\frac{1}{\omega^2 + (\xi - \gamma)^2} \, e^{i \omega t} \right\} \nonumber \\
	&
	= \frac{\gamma}{\pi} Re \Biggl\{ 
	\pi i \coth \! \left[ \frac{i \beta \hbar (\xi - \gamma)}{2} \right] \!
	e^{-(\xi - \gamma) t} 
	+ 2 \coth(1) \sin \! \left(\frac{2t}{\beta \hbar} \right) \! \beta \hbar (\xi - \gamma)
         \nonumber \\
	& \qquad \qquad \qquad \qquad  \qquad \qquad
	\vphantom{\left(\frac{2t}{\beta \hbar} \right)}
	+ \mathcal{O} \! \left(\beta \hbar \xi \right)^2 \Biggr\}
	\nonumber \\[.1in]
	\begin{split}
	&= i \gamma \coth \! \left[ \frac{i \beta \hbar (\xi - \gamma)}{2} \right] \!
	e^{-(\xi - \gamma)t} + \frac{2 \gamma}{\pi} \coth(1) 
	\sin \! \left(\frac{2t}{\beta \hbar} \right) \!
	\beta \hbar(\xi - \gamma) \\
	& \qquad \qquad \qquad \qquad  \qquad \qquad
	+ \gamma \, \mathcal{O} \! \left( \beta \hbar \xi \right)^2 .
	\end{split}
\end{align}
Arriving at the third equality requires the expression for the integral over ${\cal C}_T$ given by \eqref{A11}.  Similar to the previous case, equation \eqref{A17} and its $t=0$ form are used to obtain the $g_\xi(\omega)$ integral in \eqref{aRIntSimpxi1},
\begin{align} \label{A18}
	\frac{\gamma}{\pi} \! \int \limits_{-2/\beta \hbar}^{2/\beta \hbar} \! &d \omega
	\, \omega \, g_\xi(\omega) \coth \! \left( \frac{\beta \hbar \omega}{2} \right) \nonumber \\
	&= i \gamma \coth \! \left[\frac{i \beta \hbar (\xi - \gamma)}{2} \right] 
	\sinh\left[ (\xi - \gamma) t \right]
	- \frac{2 \gamma}{\pi} \coth(1) \sin \! \left( \frac{2t}{\beta \hbar} \right) 
	\! \beta \hbar (\xi - \gamma) \nonumber \\
	& \qquad \qquad
	+ \gamma \sinh^2 \left[ \frac{1}{2} (\xi - \gamma) t \right] \mathcal{O}(\beta \hbar \xi)^2 .
\end{align}
Equations \eqref{A16} and \eqref{A18} express the integrals over $\omega$ in terms of the required parameters.  Inserting these results back into \eqref{aRIntSimpxi1} gives
\begin{align} \label{aRIntSimpxiA3}
	&\frac{\hbar}{m} \int \limits_0^t y_{cl}(t') \int \limits_0^t a_R(t'-t'') \, y_{cl}(t'') \, dt'' dt' \nonumber \\ 
	&= -\frac{1}{2} \frac{1}{\sinh^2 (\xi t)} \bigl\{ y_f^2 \, e^{-\gamma t} \left( \zeta 
	- \varepsilon_+ e^{\xi t} - \varepsilon_- e^{-\xi t} \right)
	-2 y_f y_i \left[ \zeta \cosh(\xi t) - \varepsilon_+ - \varepsilon_- \right] \nonumber \\
	& \qquad \qquad \qquad \, \,
	+ y_i^2 \, e^{\gamma t} \left( \zeta - \varepsilon_+ e^{-\xi t} - \varepsilon_-e^{\xi t} \right)
	\bigr\} ,
\end{align}
where
\begin{subequations}
     \begin{align}
          \zeta &= i(\xi + \gamma) \coth \! \left[ \frac{i \beta \hbar (\xi + \gamma)}{2} \right] \!
          \sinh (\gamma t) + i(\xi - \gamma) \coth \! \left[ \frac{i \beta \hbar (\xi - \gamma)}{2} \right] \!
          \sinh (\gamma t) \nonumber \\
          & \quad 
          + \frac{8 \gamma}{\pi} \sinh(\xi t) \mathcal{O} (\beta \hbar \xi) \\[.15in]
          \varepsilon_\pm &= i \gamma \coth \! \left[ \frac{i \beta \hbar (\xi \pm \gamma)}{2} \right] \!
          \sinh \left[(\xi \pm \gamma) t)\right] 
          + \frac{2 \gamma}{\pi} \coth (1) \sin \! \left(\frac{2t}{\beta \hbar} \right)
          \beta \hbar (\gamma \pm \xi) \nonumber \\
          & \quad
          + \gamma \sinh^2\left[ \frac{1}{2} (\xi \pm \gamma) t \right] \mathcal{O} (\beta \hbar \xi)^2 .
     \end{align}
\end{subequations}
The terms of opposite parity in $\xi$, denoted by $\{ \xi \rightarrow -\xi \}$, have been explicitly included in $\varepsilon_\pm$ and in $\zeta$.  For times larger than or equal to the thermalization time, the limit $\gamma t \rightarrow \infty$ is applicable for which the dominant terms in $\zeta$ and $\varepsilon_\pm$ are
\begin{subequations} \label{A21}
     \begin{align}
          \zeta &\xrightarrow[\gamma t \rightarrow \infty]{} \frac{4 \gamma}{\pi} \, e^{\xi t} \,
          \mathcal{O} (\beta \hbar \xi) \\
          \varepsilon_\pm &\xrightarrow[\gamma t \rightarrow \infty]{} \frac{\gamma}{2} \,
          e^{(\xi \pm \gamma) t} \Biggl\{ i \coth \! \left[ \frac{i \beta \hbar (\xi \pm \gamma)}{2} \right]
          + \mathcal{O} (\beta \hbar \xi)^2 \Biggr\} .
     \end{align}
\end{subequations} 
Finally, expressions \eqref{A21} are used to obtain the thermalization time limit of \eqref{aRIntSimpxiA3} 
\be \label{aRIntSimpxiA4}
	\frac{\hbar}{m} \int \limits_0^t y_{cl}(t') \int \limits_0^t a_R(t'-t'') \, y_{cl}(t'') \, dt'' dt' 
	\xrightarrow[\gamma t \rightarrow \infty]{}  i \gamma \coth \! \left(\frac{i \beta \hbar \xi}{2} \right)
	 \! (y_f^2 + y_i^2 ) .
\ee
In addition to dropping terms of order $\beta \hbar \xi$ and higher, it has been assumed that $\gamma \ll \xi$, consistent with the severely underdamped case. \\





\begin{thebibliography}{00}

\bibitem{Esquinazi}
T. Scheike, W. B\"{o}hlmann, P. Esquinazi, J. Barzola-Quiquia, A. Ballestar, A. Setzer, 
Adv. Mater. doi: 10.1002/adma.201202219 (2012).

\bibitem{EngelFleming}
G. S. Engel, T. R. Calhoun, E. L. Read, T.-K. Ahn, T. Man\v{c}al, Y.-C. Cheng, R. E. Blankenship, G. R. Fleming, 
Nature 446 (2007) 782-786.

\bibitem{PanitEngel}
G. Panitchayangkoon, D. Hayes, K. A. Fransted, J. R. Caram, E. Harel, J. Wen, R. E. Blankenship, G. S. Engel, Proc. Nat. Acad. Sci. USA 107 (2010) 12766-12770.

\bibitem{CaldeiraLeggett}
A. O. Caldeira, A. J. Leggett, Physica 121A (1983) 587-616.

\bibitem{FeynmanVernon}
R. P. Feynman, F. L. Vernon, Ann. Phys. 24 (1963) 118-173.

\bibitem{FeynmanHibbs}
R. P. Feynman, A. R. Hibbs, D. F. Styer, Quantum Mechanics and Path Integrals, Emended ed., Dover, New York, 2010.

\bibitem{Rajeev}
This result is similar to that obtained for an isolated quantum oscillator where complex coordinates along with a complex-valued hamiltonian are introduced at the outset.  See S. G. Rajeev, Ann. Phys. 322 (2007) 1541-1555. 


\bibitem{Lloyd}
P. Rebentrost, M. Mohseni, I. Kassal, S. Lloyd, A. Aspuru-Guzik, New J. Phys. 11 (2009) 033003.

\bibitem{Pleino}
M. B. Plenio, S. F. Huelga, New J. Phys. 10 (2008) 113019.



\end{thebibliography}



\end{document}